\documentclass[journal]{IEEEtran}

\usepackage[]{graphicx}
\graphicspath{{Figures/}}
\usepackage{caption}
\usepackage{subfig} %should come after caption package if given, can't be used with caption2, defines subfloat
\usepackage{epsfig}
\usepackage{cite}
\usepackage{color}
\usepackage{balance}
\usepackage{amsmath}
\usepackage{nicefrac}
\usepackage{amssymb}
\usepackage{amsfonts} % for the \checkmark command 
\usepackage{bm}
\usepackage{graphicx}
\usepackage{pdfpages}
\usepackage[T1]{fontenc} % For less than and greater than sign
\usepackage{array}
\usepackage[hyphens,spaces,obeyspaces]{url}
\usepackage{color}
\usepackage{wrapfig}
\usepackage{lipsum}
\usepackage[hidelinks]{hyperref}
\usepackage{multirow}
\usepackage{tabularx}
\usepackage{tikz}
\usepackage{lscape}
\usepackage{xcolor}
\usepackage{censor}
\usepackage{stfloats}% To put table and figures at the bottom of page by force.

\DeclareMathOperator*{\argmin}{argmin}

\usepackage[linesnumbered,ruled,vlined]{algorithm2e}
\SetKwComment{Comment}{$\triangleright$\ }{}
\let\oldnl\nl% Store \nl in \oldnl
\newcommand{\nonl}{\renewcommand{\nl}{\let\nl\oldnl}}% Remove line number for one line
\newcommand{\vars}{\texttt}

\usepackage{comment}
\usepackage[compact]{titlesec}         % you need this package
\DeclareCaptionLabelSeparator{periodspace}{.\quad}

\usepackage{nomencl}
\makenomenclature

\begin{document}
 \pagenumbering{gobble} 

%\title{Clustering Residential Electric Load\\ Demand Profiles}
\title{Validating Clustering Frameworks for \\Electric Load Demand Profiles}

\author{
Mayank~Jain,
Tarek~AlSkaif~\IEEEmembership{Member,~IEEE}
and~Soumyabrata~Dev~\IEEEmembership{Member,~IEEE}% <-this % stops a space
\thanks{
Manuscript received August 26, 2020; revised December 02, 2020; and accepted February 08, 2021.
}

\thanks{
M.\ Jain and S.\ Dev are with ADAPT SFI Research Centre, University College Dublin, Ireland (e-mail: mayank.jain1@ucdconnect.ie, soumyabrata.dev@ucd.ie). T.\ AlSkaif is with Information Technology Group, Wageningen University and Research, Wageningen, The Netherlands (e-mail: tarek.alskaif@wur.nl).
}% <-this % stops a space
\thanks{
Send correspondence to S.\ Dev, E-mail: soumyabrata.dev@ucd.ie.}% <-this % stops a space
\thanks{
The ADAPT Centre for Digital Content Technology is funded under the SFI Research Centres Programme (Grant 13/RC/2106) and is co-funded under the European Regional Development Fund.}
}

\markboth{IEEE Transactions on Industrial Informatics,~Vol.~XX, No.~XX, XX~2021}%
{Shell \MakeLowercase{\textit{et al.}}: Bare Demo of IEEEtran.cls for IEEE Journals}



\maketitle

\begin{abstract}
Large-scale deployment of smart meters has made it possible to collect sufficient and high-resolution data of residential electric demand profiles. Clustering analysis of these profiles is important to further analyze and comment on electricity consumption patterns. Although many clustering techniques have been proposed in the literature over the years, it is often noticed that different techniques fit best for different datasets. To identify the most suitable technique, standard clustering validity indices are often used. These indices focus primarily on the intrinsic characteristics of the clustering results. Moreover, different indices often give conflicting recommendations which can only be clarified with heuristics about the dataset and/or the expected cluster structures -- information that is rarely available in practical situations. This paper presents a novel scheme to validate and compare the clustering results objectively. Additionally, the proposed scheme considers all the steps prior to the clustering algorithm, including the pre-processing and dimensionality reduction steps, in order to provide recommendations over the complete framework. Accordingly, the proposed strategy is shown to provide better, unbiased, and uniform recommendations as compared to the standard Clustering Validity Indices.
\end{abstract}


\vspace{1em}
\IEEEpeerreviewmaketitle
\section*{Nomenclature}
\addcontentsline{toc}{section}{Nomenclature}
\begin{IEEEdescription}[\IEEEusemathlabelsep\IEEEsetlabelwidth{7.7em}]
% Nomenclature
%\nomenclature{CVI(-s)}{Clustering Validity Index(-ices)}
\item[CVI(-s)]{Clustering Validity Index(-ices)}
\item[PCA]{Principal Component Analysis}
\item[FA]{Feature Agglomeration}
\item[KMC]{K-Means Clustering}
\item[SC]{Spectral Clustering}
\item[AC]{Agglomerative Clustering}
\item[FCM]{Fuzzy C-Means Clustering}
\item[$n$]{Number of households}
\item[$r$]{Hourly resolution of smart meters}
\item[$d \equiv \nicefrac{24}{r}$]{Original dimensionality of consumption profiles}
\item[$d'$]{Optimal number of reduced dimensions}
\item[$k$]{Optimal number of clusters}
\item[feature]{Vector by which load consumption profile of an household is being represented at a time}
\end{IEEEdescription}

\printnomenclature

\vspace{1em}
\section{Introduction}
\label{sec:intro}
%~\cite{singh2017mining} as optional behavioural analysis paper
\IEEEPARstart{P}{ossessing} the capability to measure and record the electric power consumption data of a household, smart-meters have been immensely helpful towards the realization of smart grids. The captured data is typically communicated to the energy suppliers at high temporal resolutions, enabling them to perform data analytics in order to better understand the consumption behavior of their customers~\cite{motlagh2015}. Gaining insights into customers' electric load demand profiles will not only help in preparing a better energy generation plan for the smart grids but will also enable the energy suppliers to optimize electricity tariff structures~\cite{gungor2012survey}. Clustering these profiles is an important task to gain such insights.

The process of clustering residential load demand profiles has been extensively studied in prior literature~\cite{williams2013,hino2013,zhou2017,yilmaz2019comparison,al2016,lin2017,waczowicz2015,motlagh2015,mcloughlin2015,yildiz2018,viegas2015,yang2019amodel}. While some just attempted to cluster the electric demand profiles of households~\cite{williams2013,hino2013,zhou2017,yilmaz2019comparison,al2016}, others aimed to use the clustering results for further analysis~\cite{lin2017,waczowicz2015,motlagh2015,mcloughlin2015,yildiz2018,viegas2015, yang2019amodel}. In every case, a crucial task is to perform the best possible clustering. However, defining the \textit{best} is a challenge faced by all prior studies. This is because, for a given dataset, different clustering algorithms produce different results; and any single clustering algorithm does not work equally good for all datasets~\cite{mccarthy2019exact}. This means that there is a strong requirement of an approach (or algorithm) which can compare the results of many different clustering algorithms and give a recommendation for one based on its performance on a given dataset.

When ground truth cluster labels are available, it's easy to calculate the accuracy of the clustering results and consecutively identify the best clustering approach. However, in the case of unsupervised clustering tasks with completely unlabeled data, accuracy can not be computed directly. To resolve this issue, literature offers many indices to compare the quality of clustering results~\cite{rousseeuw1987, calinski1974, dunn1973, davies1979, xie1991}. These indices are commonly referred to as the Clustering Validity Indices (CVIs). However, these indices only focus on the input and output of the employed clustering algorithms and do not consider the pre-clustering steps. Adding onto the fact that different indices look at different intrinsic aspects of the clustering results, it is often difficult to unanimously identify the best clustering results which can be used for further analysis in order to gain insights on the electricity consumption pattern of the consumers.

\subsection{Related Work}\label{sec:litReview}
The problem of identifying the best clustering approach is not new in the domain of clustering residential load demand profiles. Note that a clustering approach is defined by the steps of the clustering framework, the algorithm used for each step, and the hyperparameter settings of those algorithms (refer to Section~\ref{sec:clusteringFramework} for more details). Changing any of these will lead to a different clustering approach.

In an elaborate discussion on clustering residential electricity consumption profiles, Williams J.~\cite{williams2013} considers multiple clustering algorithms, with and without feature selection and dimensionality reduction techniques. In the absence of true labels, the validation of clustering results was done by using different CVIs. Those indices were further used to determine the optimal number of clusters individually for the considered clustering algorithms. Not surprisingly, different indices provide different recommendations for the clustering algorithms with the optimal number of clusters (hyperparameter setting) ranging from $3$ to $20$. Finally, performing a closer subjective inspection of the generated clusters led the author to determine the best clustering approach for the used dataset. In a similar situation, Viegas~\textit{et~al.}~\cite{viegas2015} opted to use the majority vote on the recommendations made by different CVIs to select the optimal number of clusters for the only clustering approach which was explored in the paper.

Some other studies simply chose to navigate through this difficulty by selectively reducing the exploration domain. McLoughlin~\textit{et~al.}~\cite{mcloughlin2015} used only the Davies-Bouldin index~\cite{davies1979}, both for choosing the best clustering algorithm and the optimal number of clusters. While considering different aspects of analyzing the raw daily consumption data, Yilmaz~\textit{et~al.}~\cite{yilmaz2019comparison} primarily used the K-Means algorithm for clustering and the Silhouette index~\cite{rousseeuw1987} for identifying the optimal number of clusters. Zhou~\textit{et~al.}~\cite{zhou2017}, Hino~\textit{et~al.}~\cite{hino2013}, and Yang~\textit{et~al.}~\cite{yang2019amodel} also chose to work with only one clustering algorithm. While Zhou~\textit{et~al.} made a comparison between $4$ and $6$ clusters, Hino~\textit{et~al.} and Yang~\textit{et~al.} used gap-statistics~\cite{tibshirani2001} and histogram of distances~\cite{yang2019amodel}, respectively, to choose hyperparameter settings without any further validation. Al-Otabi~\textit{et~al.}~\cite{al2016} and Waczowicz~\textit{et~al.}~\cite{waczowicz2015} specified their own CVI and used it to obtain their results. Whereas, %Gouveia \& Seixas~\cite{gouveia2016}, Rhodes~\textit{et~al.}~\cite{rhodes2014}, and
Motlagh~\textit{et~al.}~\cite{motlagh2015} not only worked with just one clustering approach but also heuristically opted for a fixed number of clusters.

From the numerous examples that are discussed above, it can be noted that the CVIs often provide different recommendations over the considered clustering approaches leading the researchers to rely mainly on less standard and more subjective approaches like manual inspection, heuristics, majority voting or simply considering limited number of clustering approaches and/or CVIs. This calls for a more robust and standard objective validation strategy to holistically judge the clustering results.

\subsection{Paper Contribution}
%\textit{Add a paragraph here!}
The primary objective of this paper is to identify an optimal approach to efficiently cluster households based on their daily load consumption profiles. To this end, a novel objective validation strategy has been proposed to compare the output of different clustering approaches. In order to understand the proposed algorithm and its significance in a better way, the paper formalizes the definition of a clustering approach by presenting a generalized $2$-step clustering framework~\cite{jain2020clustering}. Furthermore, a comparative analysis between the recommendations made by the proposed strategy and the ones made by standard CVIs is done to determine the advantages of the proposed strategy. These contributions can be summarized as follows\footnote{In the spirit of reproducible research, the code to reproduce the results in this paper is shared at \url{https://github.com/jain15mayank/validating-clustering-frameworks}.}:
\begin{itemize}
    \item Presentation of a generalized $2$-step clustering framework;
    \item Proposal of a novel objective validation strategy to compare different clustering frameworks; and
    \item Comparison of the proposed strategy with standard CVIs.
\end{itemize}

The rest of the paper is structured as follows. Section~\ref{sec:dataset} describes the dataset used in this study, along with the pre-processing steps. Section~\ref{sec:clusteringFramework} presents the generalized $2$-step clustering framework along with the algorithms that are discussed for each step of the clustering framework. Section~\ref{sec:valCFs} defines the proposed objective validation strategy and states the standard CVIs which were used to perform the comparative analysis. The results of this study are presented and discussed in Section~\ref{sec:results}, and the paper is concluded in Section~\ref{sec:conclusion}.

\section{System Design: Dataset \& Pre-processing}
\label{sec:dataset}

This research considers realistic electricity consumption profiles of actual households who were participants in a European pilot project for sustainable energy management in residential communities. Hourly electricity consumption data of each household was captured by its smart meter and forwarded to a web-based participatory platform in an automated manner. The data collection period was from January 2018 up until March 2019. In this pilot, $27$ households, situated in Amsterdam, the Netherlands, provided consent for their data to be accessed from the web platform and used for scientific research purposes. Since the project asks for voluntary participation of the households, it is called as the ``PARTicipatory platform for sustainable ENergy management'' (PARENT)~\cite{PARENT}. The final aim of the project was to analyze this data and provide recommendations for different stakeholders, namely: \textit{i)} for households to reduce their electricity consumption (e.g., by providing instant feedback and periodic messages), and \textit{ii)} for energy suppliers to gain better insights about their customers. This paper contributes to the second objective of this project by proposing an validation strategy to compare the clustering results of households' electric load demand profiles objectively.

In this paper, we do not consider the impact of seasonality on the electricity consumption of households due to the paucity of data spanning multiple years. This assumption can be relaxed once more data is available from the pilot project. However, it is worth noting that this assumption does not create any impact, neither on the clustering framework nor on the proposed objective validation strategy. This is because data for each season can be considered separately and hence can be managed in the pre-processing steps (e.g. independently clustering households for different seasons).

\begin{figure}[!htb]
  \begin{center}
  \includegraphics[width=.95\linewidth]{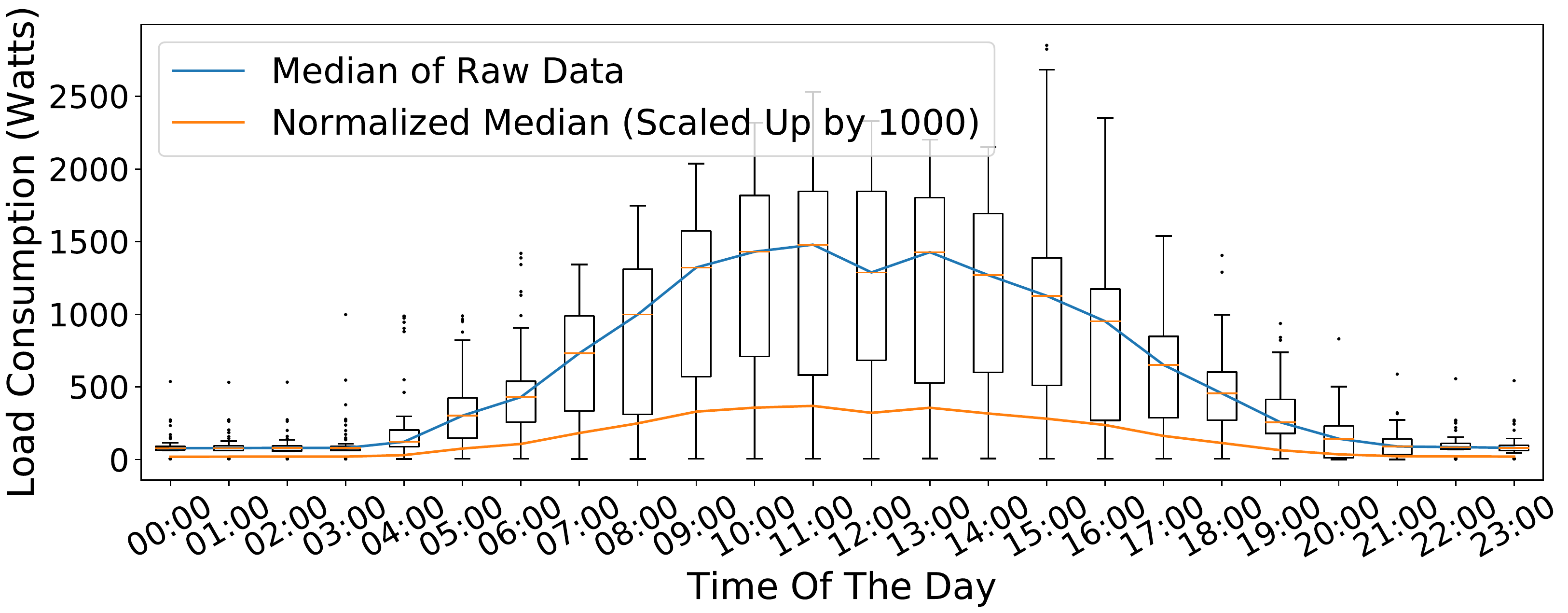}\label{fig:box-plot1}
  %\newline
  \includegraphics[width=.95\linewidth]{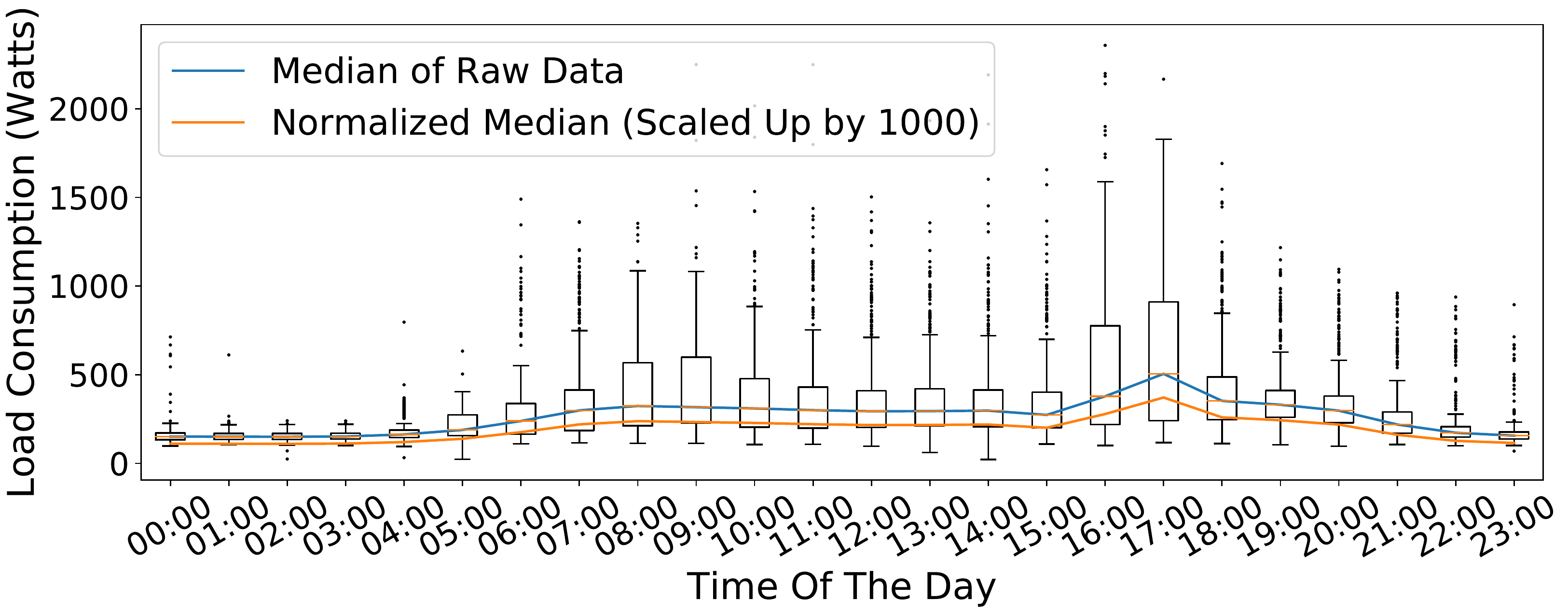}\label{fig:box-plot2}
  \end{center}
\caption{Box-plots of daily consumption pattern for two different households in the dataset. It also shows the median and the scaled up version of normalized median.}
\label{fig:BoxPlots}
\end{figure}

Figure~\ref{fig:BoxPlots} shows representative box-plots for $2$ distinct households from the considered dataset. These box-plots demonstrate the households' electric load consumption patterns across the different hours of the day. It can be further observed that the pattern of consumption varies significantly for the two households. Median consumption patterns for the households are also shown in these plots. Since the median consumption patterns correctly approximate the electricity consumption trends of the households, these patterns are used to define the daily consumption profiles for the households. Lastly, to identify the consumption trends across the different households while ignoring the scale of consumption, $\ell_2$-normalization is performed on the median profiles obtained for each household. These steps are illustrated in Algorithm~\ref{algo:preProcess}. Values of the variables, specific to the dataset which is used in this study, are written in braces alongside the variable declaration steps of Algorithm~\ref{algo:preProcess}.

\begin{algorithm}
\caption{Pre-processing the input dataset}
\label{algo:preProcess}
%\DontPrintSemicolon % Some LaTeX compilers require you to use \dontprintsemicolon instead
$\vars{n} \gets \text{number of households } (= 27)$\\
$\vars{r} \gets \text{hourly resolution of original data } (= 1)$\\
$\vars{d} \gets 24/r$ \text{(dimensionality)}\\
$\bm{M_{n \times d}} \gets$ median consumption of each household stored row-wise (load demand profiles)\\
$\bm{M'_{n \times d}} \gets \ell_2\text{-Normalization(}\bm{M}\text{, row-wise)}$\\
\textbf{return} $\bm{M'}$
\end{algorithm}

\section{Clustering Framework}
\label{sec:clusteringFramework}

Dimensionality reduction techniques have been widely used before applying clustering algorithms over load demand profiles in high dimensions~\cite{williams2013,lin2017,yildiz2018,al2016}. Not only do they significantly reduce the computational complexity but they also help avoid overfitting and achieve more meaningful clustering results~\cite{al2016}. This can be further confirmed with the results presented in TABLE~\ref{table:dimReduceUtilityVerification}, where different clustering frameworks were timed. We can note that adding the dimensionality reduction step generally reduces the overall time complexity of the clustering framework. A better time complexity can further be correlated to better scalability of the clustering framework, thereby further emphasizing on the importance of dimensionality reduction techniques.

\begin{table}[htb!]
\small
\centering
\caption{TIME TAKEN (in $ms$) TO PERFORM CLUSTERING\\(AVERAGED OVER $100$ TRIALS FOR $5$ CLUSTERS)} %WITH AND WITHOUT DIMENSIONALITY REDUCTION TECHNIQUES\\(AVERAGED OVER $100$ TRIALS FOR $5$ CLUSTERS)}
\begin{tabular}{p{3.75cm}||c|c|c|c}
\hline
Dimensionality Reduction & KMC & SC & AC & FCM \\ 
\hline\hline
No Reduction & \textbf{18.35} & \textbf{17.83} & \textbf{0.34} & \textbf{3.10}\\
PCA & 17.75 & 17.58 & \textbf{0.34} & 2.81\\
FA & 17.47 & 17.39 & 0.33 & 2.44\\
\hline
\end{tabular}
%\vspace{-0.5cm}
\label{table:dimReduceUtilityVerification}
\end{table}

In the case of clustering load demand profiles, it is more important for the clustering algorithms to focus on the crucial characteristics of the profile, like peak times for increased consumption, and trend~\cite{lin2017}. Properly reducing the dimensions helps analyze these distinctive characteristics while removing unnecessary or irrelevant features, making it easier for the clustering algorithms to focus on them.

\begin{figure}[htb]%{.47\textwidth}
  \centering
  \includegraphics[width=.925\linewidth]{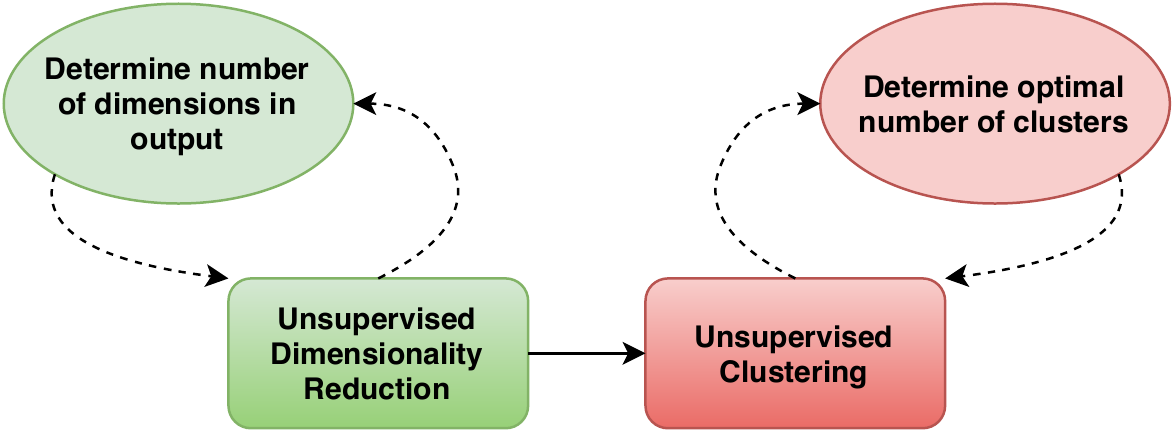}  
  \caption{Generalized Clustering Framework}
  \label{fig:cluster_framework}
\end{figure}

Both dimensionality reduction and the clustering algorithms need to be optimized before the final implementation. For instance, determining the optimal number of reduced dimensions is a crucial hyperparameter for most dimensionality reduction techniques, whereas determining the optimal number of clusters is the key hyperparameter in the clustering step. Tuning and determining the optimal values of these hyperparameters can be considered as the intermediate step inherent to the dimensionality reduction and the clustering steps, respectively. The resulting generalized clustering framework is presented in Fig.~\ref{fig:cluster_framework}. Additionally, since this paper deals with the task of clustering residential load demand profiles, a hard limit of $10$ as the maximum number of clusters is set. This is in accordance with the recommendations made by industry professionals for practical relevance, as suggested by Al-Otabi~\textit{et~al.}~\cite{al2016}.

\subsection{Unsupervised Dimensionality Reduction}\label{sec:dimReduce}
\begin{figure}[htb]%{.47\textwidth}
  \centering
  \includegraphics[width=.925\linewidth]{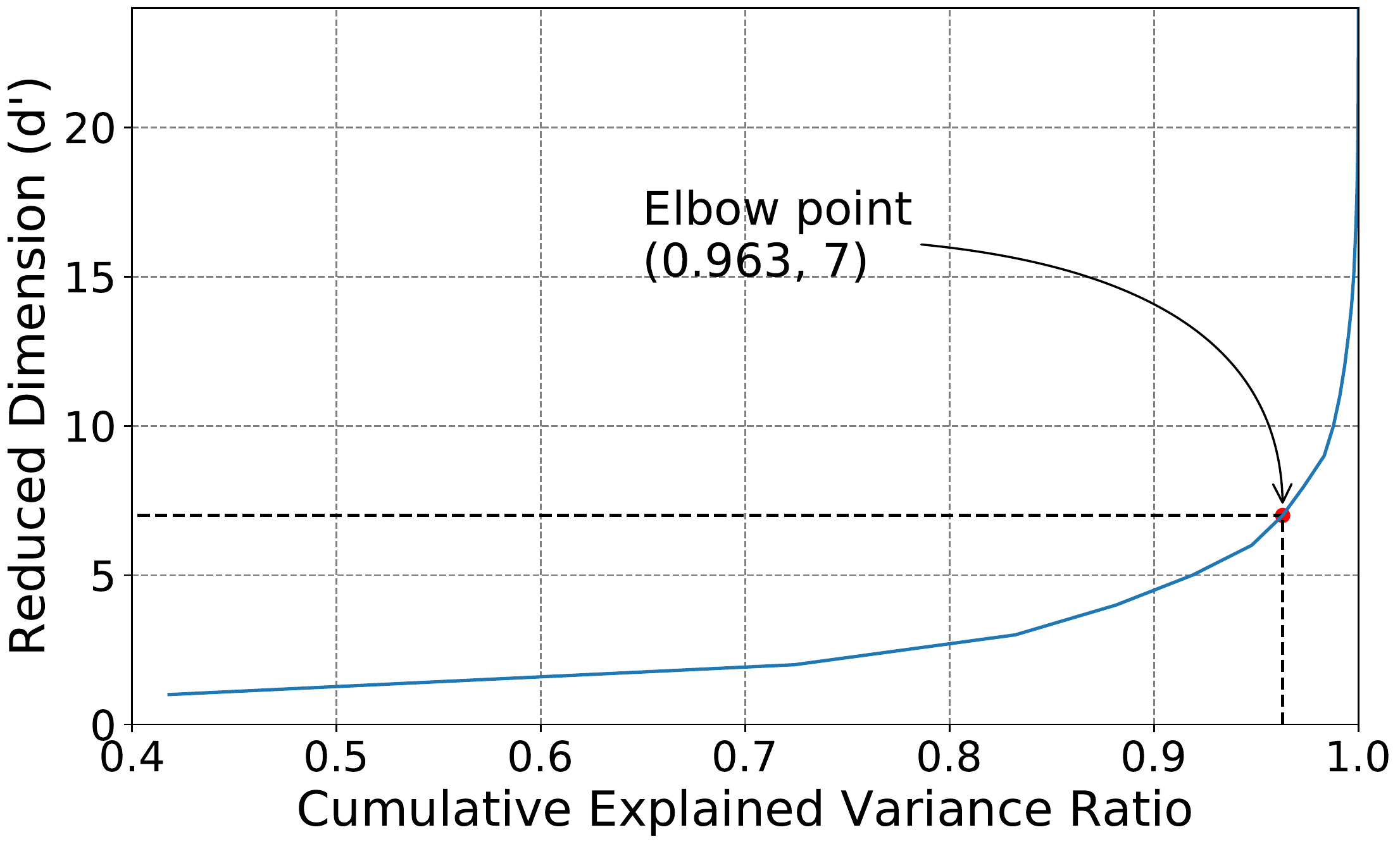}  
  \caption{CEVR vs. $d'$ for PCA hyperparameter setting.}
  \label{fig:elbow-pca}
\end{figure}

%~\cite{geladi1989,petroni2000,sophian2003} instead of ~\cite{wold1987,yildiz2018}
\textbf{Principal Component Analysis} (PCA) is one of the most widely applied algorithms for this task~\cite{wold1987,yildiz2018}. Consider a dataset of $n$ households where every household has features in $d$ dimensions that are representative of their load demand profiles. This can then be represented by a matrix $A$ of size $n \times d$. The next step is to compute the covariance matrix $\Psi_{A}$. This is followed by the computation of eigenvalues ($\bar{\lambda}$) and corresponding eigenvectors ($\bar{\boldsymbol{u}}_{d\times1}$) of the covariance matrix. After sorting the eigenvalues in descending order ($\bar{\lambda}_{sorted}$), the corresponding first $d'_{PCA}$ eigenvectors are picked to create a transformation matrix $W$ of size $d \times d'_{PCA}$. The final step is just to multiply $A$ with $W$ to obtain $A'$ of size $n \times d'_{PCA}$, thereby reducing the dimensions (see equation \ref{eq:PCAfinal}).
\begin{equation}
    A'_{n \times d'_{PCA}} = A_{n \times d} \cdot W_{d \times d'_{PCA}} 
    \label{eq:PCAfinal}
\end{equation}
To determine the value of $d'_{PCA}$ (i.e., the reduced number of dimensions), the property of the eigenvalues ($\bar{\lambda}$) to represent explained variance is utilized~\cite{wold1987}. In this regard, the cumulative explained variance ratio ($CEVR$) of the sorted eigenvalues was computed as described in equation \ref{eq:CEVR}.
\begin{equation}
    {CEVR}(\lambda_{sorted}^{i}) = \dfrac{\sum_{j=1}^{i} \lambda_{sorted}^{j}}{\sum_{k=1}^{d} \lambda_{sorted}^{k}}
    \label{eq:CEVR}
\end{equation}
This relationship is then used to create a graph between $CEVR$ and $d'_{PCA}=i$ on which elbow heuristics is performed to determine the final values of hyperparameters (\textit{cf.} Fig.~\ref{fig:elbow-pca}): $CEVR = 0.96$, and $d'_{PCA} = 7$.

Another dimensionality reduction technique that is explored in this work is \textbf{Feature Agglomeration} (FA). It is quite similar to the AC algorithm~\cite{sokal1958} (see Section \ref{sec:clusteringAlgos}) but instead of clustering the households, this works by clustering their features (or their load demand profiles). Similar to the case of AC, a distance measure is calculated between the dimensions to measure closeness. The distance measure used in this paper is based on the `Ward' linkage criterion~\cite{ward1963}, where the variance of individual features is compared. Then the two features which are closest to each other are merged. While merging, a pooling function (`arithmetic mean' in this paper) is used to merge the two features. This process is repeated until exactly one of the terminal conditions is met. These conditions are:
\begin{itemize}
    \item The remaining number of features equals the defined number of clusters
    \item The distance between the two clusters which are being merged is greater than the defined distance threshold
\end{itemize}

\begin{figure}[htb]%{.47\textwidth}
  \centering
  \includegraphics[width=.925\linewidth]{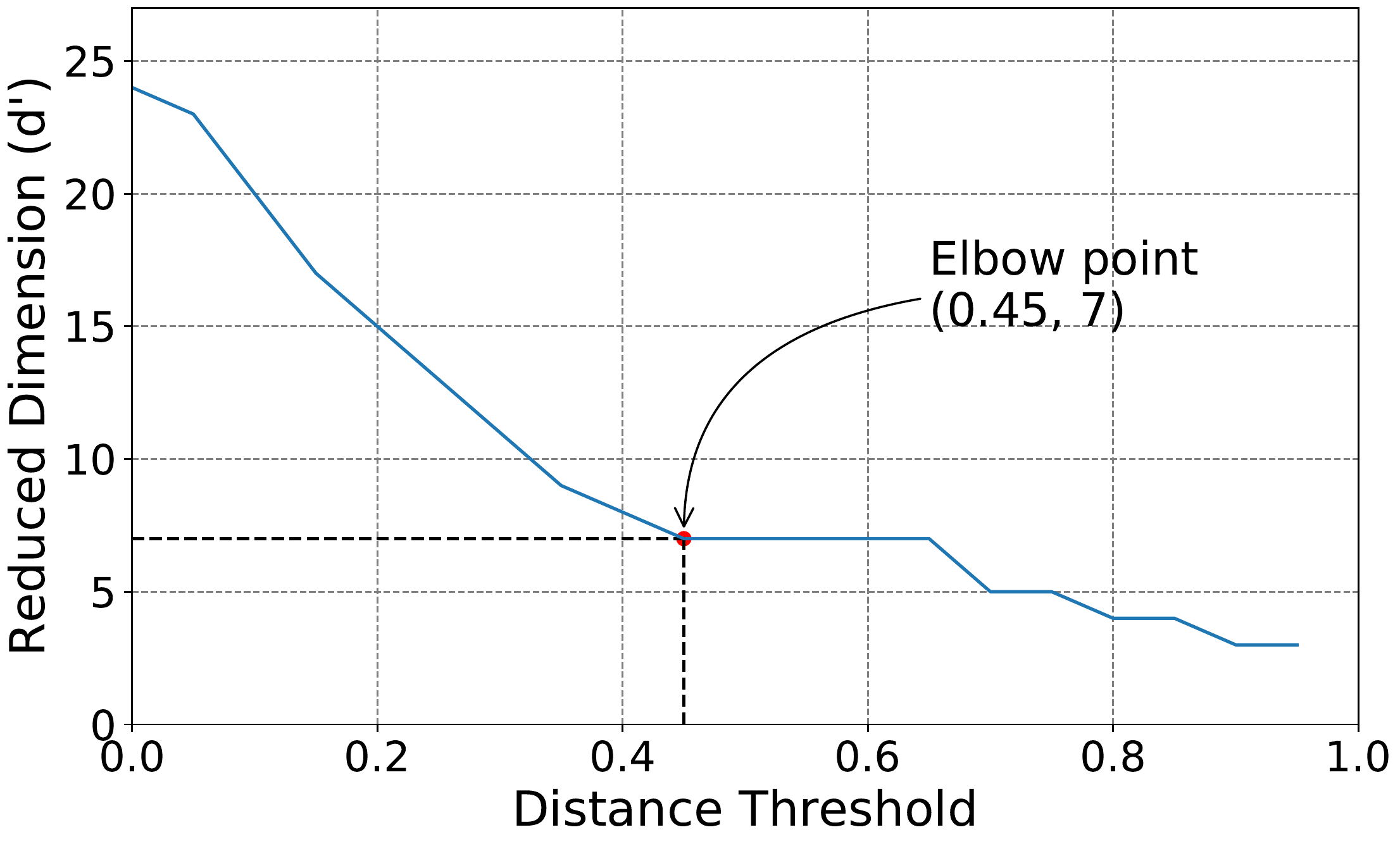}  
  \caption{$thr_{FA}$ vs. $d'$ for FA hyperparameter setting.}
  \label{fig:elbow-fa}
\end{figure}

%~\cite{mccallum2000} instead of ~\cite{sokal1958}
Similar to PCA, the inherent task is to determine the optimal number of clusters (say, $d'_{FA}$) or the distance threshold (say, $thr_{FA}$). As the value of $thr_{FA}$ increases, more features will be merged resulting in loss of information at every subsequent merge~\cite{sokal1958}. Therefore, elbow heuristics is applied yet again by plotting $d'_{FA}$ against $thr_{FA}$ (\textit{cf.} Fig.~\ref{fig:elbow-fa}) and the final values of the hyperparameters are determined: $thr_{FA} = 0.45$, and $d'_{FA} = 7$.

\subsection{Unsupervised Clustering Algorithms}\label{sec:clusteringAlgos}
\begin{figure}[htb]
\begin{center}
    \includegraphics[width=.925\linewidth]{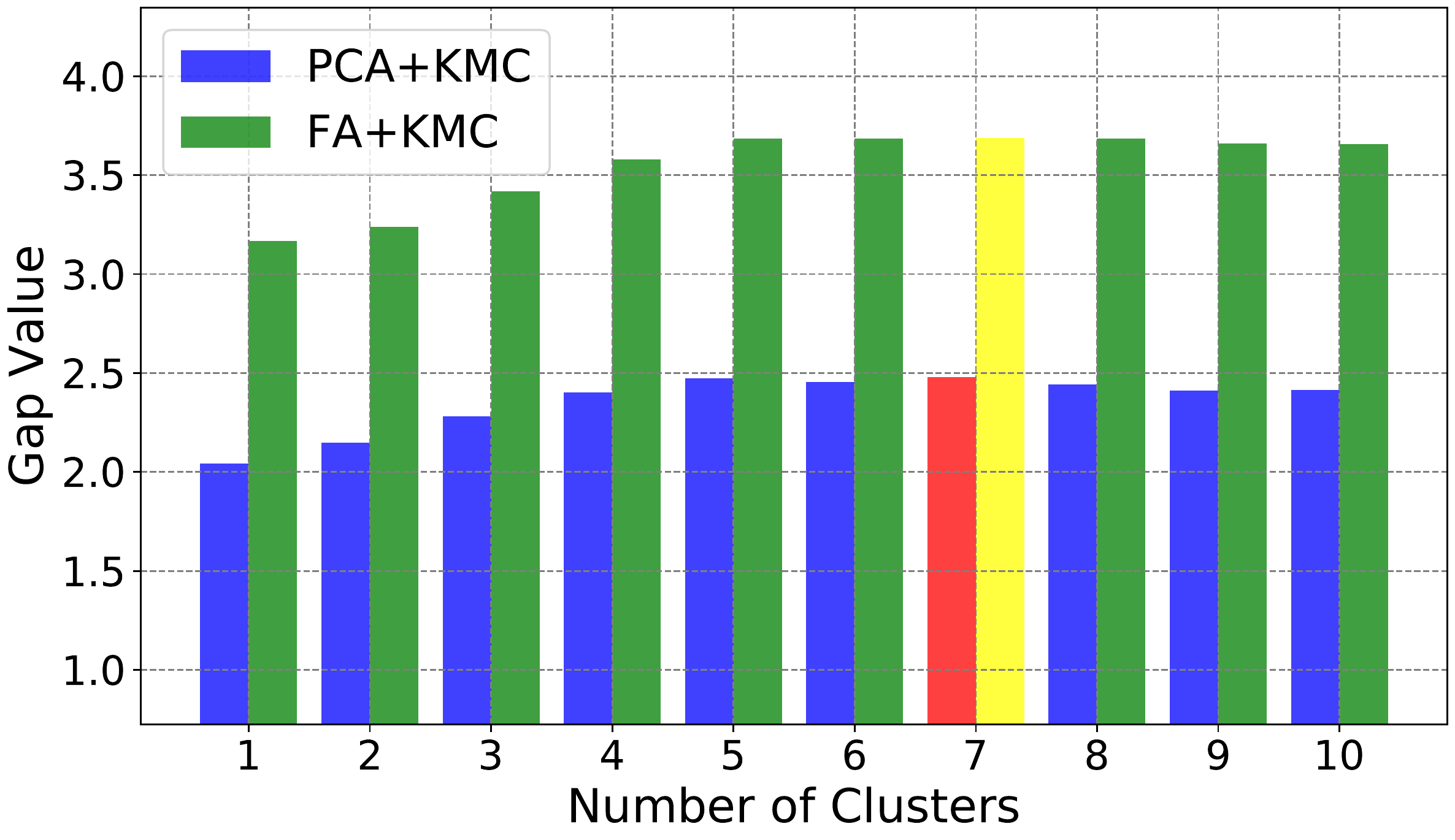}
\end{center}
\caption{Results of gap statistics as obtained for two different clustering frameworks with KMC as the clustering algorithm (RED: max gap value for PCA$+$KMC, YELLOW: max gap value for FA$+$KMC).}
\label{fig:gapKMC}
\end{figure}

One of the most popular algorithms in this category is the \textbf{k-Means Clustering} (KMC) algorithm. Given a target number of clusters to be formed (say, $k_{KMC}$), the algorithm iteratively changes the position of cluster centroids in order to find the best possible clustering. Here, the hyperparameter is $k_{KMC}$, which is calculated with the aid of gap statistics~\cite{tibshirani2001}. To find the optimal number of clusters, this process defines a statistic called the \textit{gap} for the different number of clusters and then chooses the one corresponding to the maximum value of \textit{gap}. In essence, \textit{gap} signifies how distinct is the clustering compared to a random uniform distribution of points. Since the clustering step is a successor of the dimensionality reduction step, its output depends largely on the preceding step. Hence for the two different clustering frameworks, i.e. `PCA$+$KMC' and `FA$+$KMC', the obtained optimal hyperparameter settings are: $k_{PCA+KMC} = 7$ and $k_{FA+KMC} = 7$ respectively (\textit{cf.} Fig.~\ref{fig:gapKMC}).

\begin{figure}[htb]
\begin{center}
    \includegraphics[width=.925\linewidth]{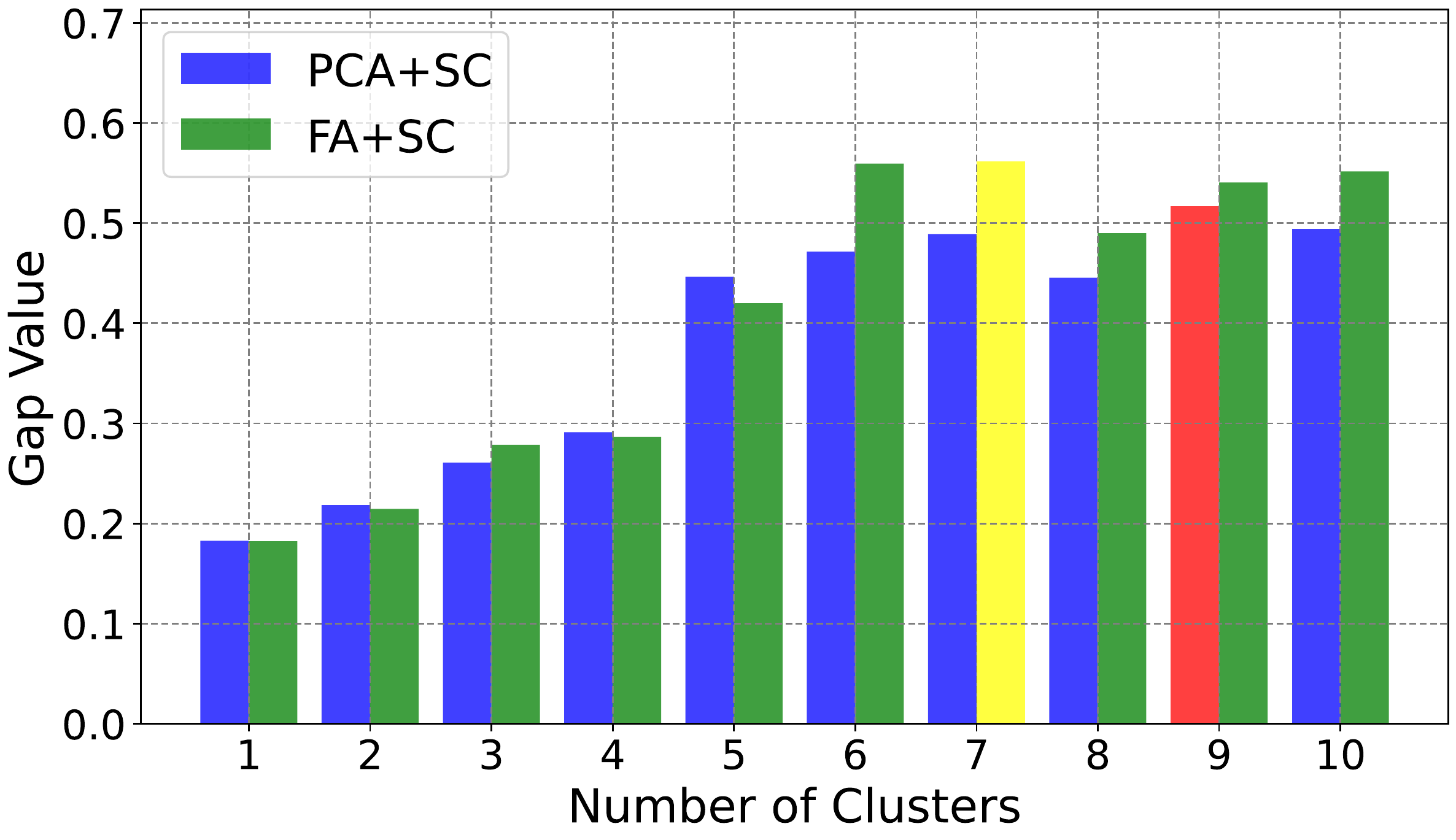}
\end{center}
\caption{Results of gap statistics as obtained for two different clustering frameworks with SC as the clustering algorithm (RED: max gap value for PCA$+$SC, YELLOW: max gap value for FA$+$SC).}
\label{fig:gapSC}
\end{figure}

A relatively less popular clustering algorithm, in the context of clustering residential load demand profiles~\cite{lin2017}, is also explored in this work, i.e. \textbf{Spectral Clustering} (SC). There are many variants of the SC algorithm~\cite{shi2000,ng2002,von2008}. This work focuses on one such variant using unnormalized Laplacian of the graph obtained by the undirected k-Nearest Neighbour algorithm~\cite{von2008}. This technique is based on the idea that the households whose demand profiles lie spectrally or spatially close to each other will fall into the same cluster. Similar to the case of KMC, the hyperparameter is to determine the optimal number of clusters. Hence, gap statistic is used once again to determine the final hyperparameter settings: $k_{PCA+SC} = 9$ and $k_{FA+SC} = 7$ (\textit{cf.} Fig.~\ref{fig:gapSC}).

\begin{figure}[htb]
\begin{center}
    \includegraphics[width=.925\linewidth]{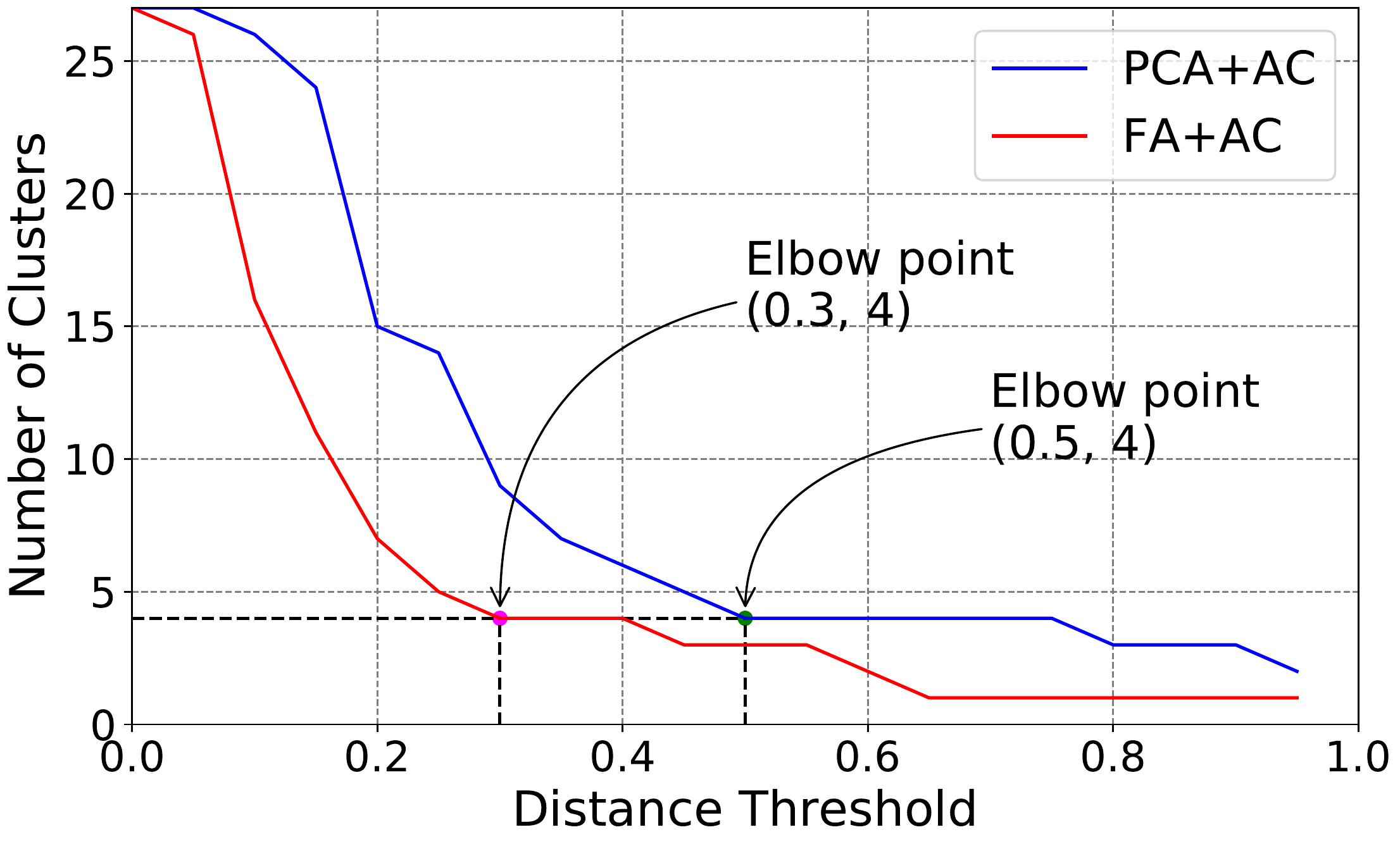}
\end{center}
\caption{Finding hyperparameter settings using elbow heuristics for $2$ different clustering frameworks with AC as the clustering algorithm, i.e. PCA$+$AC, and FA$+$AC.}
\label{fig:elbowAC}
\end{figure}

For the task of clustering residential load demand profiles, hierarchical clustering algorithms are widely used~\cite{williams2013,hino2013}. Therefore, the next algorithm explored in this category is \textbf{Agglomerative Clustering} (AC)~\cite{sokal1958}. As explained in Section~\ref{sec:dimReduce}, in the context of FA, `Ward' linkage criterion~\cite{ward1963} is used to agglomerate households into the clusters until a certain distance threshold is reached or a certain number of clusters remain. Similar to the case of FA, elbow heuristics is used to determine these hyperparameter settings: $k_{PCA+AC} = 4$ and $k_{FA+AC} = 4$ (\textit{cf.} Fig.~\ref{fig:elbowAC}).

Instead of hard (or definite) labels, \textbf{Fuzzy C-Means} (FCM) clustering algorithm identifies fuzzy labels and is the last candidate algorithm that is explored in this work. Many recent studies about clustering residential electric demand profiles use this algorithm~\cite{zhou2017,waczowicz2015,viegas2015}. It associates each household (say, $\bm{x}$) with all the clusters using a membership value  (or the degree of association). For $n$ households and $k_{FCM}$ clusters, the algorithm has $2$ hyperparameters to define a priori. One is the value of fuzzifier ($m$), while the other being the number of clusters to be generated ($k_{FCM}$).

\begin{figure}[htb]
\begin{center}
    \includegraphics[width=.925\linewidth]{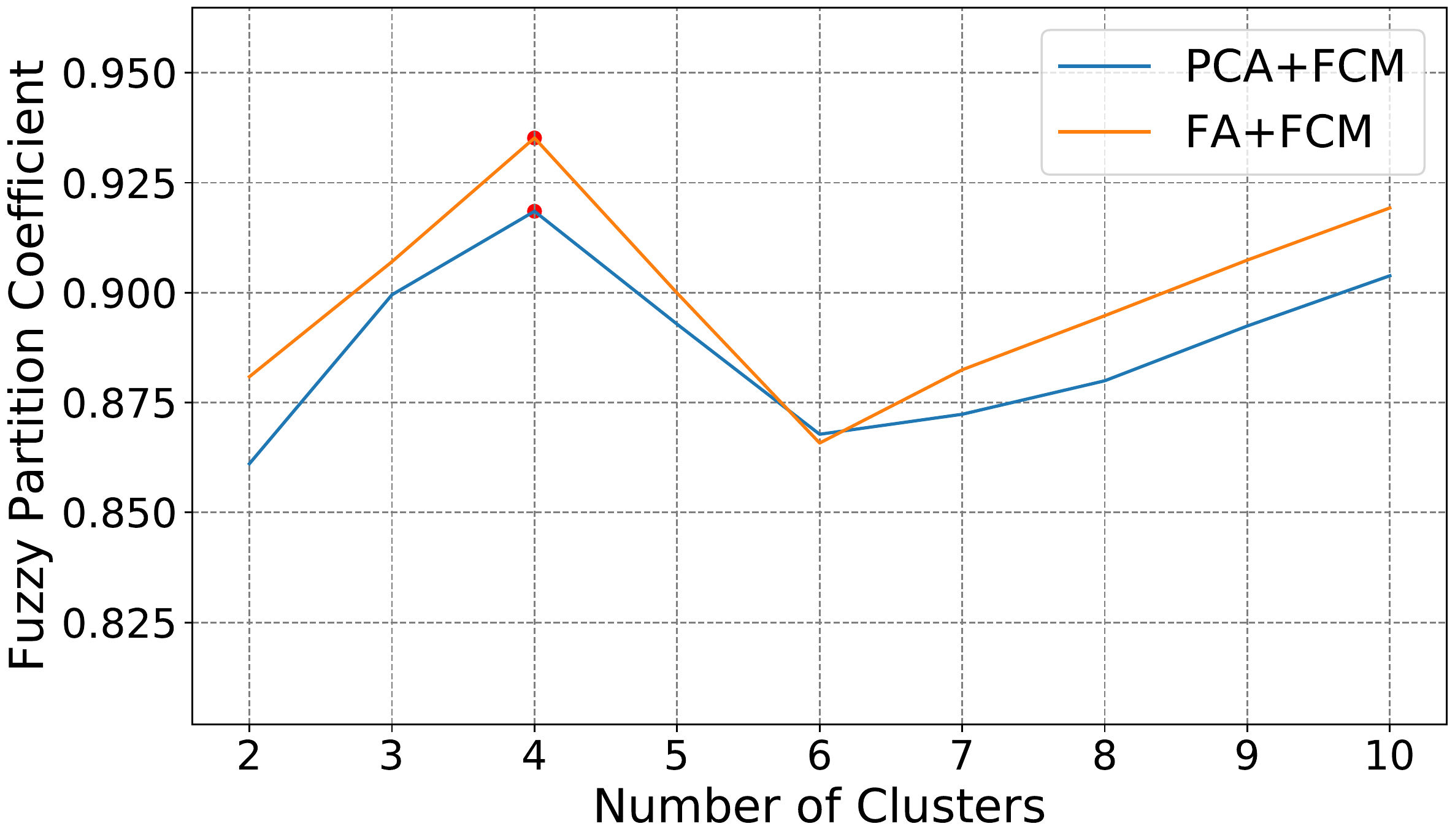}
\end{center}
\caption{Computed values of FPC for varying number of clusters as obtained for two different clustering frameworks with FCM as the clustering algorithm.}
\label{fig:FPCFCM}
\end{figure}

While optimal value of fuzzifier ($m$) was calculated using the approach defined by Demb\'{e}l\'{e} \& Kastner~\cite{dembele2003}, the optimal number of clusters was calculated by calculating the value of \textit{Dunn's fuzzy partition coefficient} (FPC)~\cite{dunn1973} for different number of clusters and choosing the value of $k_{FCM}$ where FPC is maximum (\textit{cf.} Fig.~\ref{fig:FPCFCM}). The final values of hyperparameters in this case are: $m_{PCA+FCM} = 1.3053$, $k_{PCA+FCM} = 4$, and $m_{FA+FCM} = 1.3125$, $k_{FA+FCM} = 4$.

\section{Validating Clustering Frameworks}
\label{sec:valCFs}
An even more important task is to validate and compare the results which were obtained by the different frameworks (as defined in Section~\ref{sec:clusteringFramework}), in order to pick the best clustering framework for further analysis. Most researchers have relied on the numerous CVIs for this purpose~\cite{mcloughlin2015,williams2013,zhou2017,yildiz2018,viegas2015}. In the absence of true clustering labels, these CVIs primarily focus on the intrinsic properties of the clustering results, namely, compactness and separation~\cite{liu2010}. A more comprehensive discussion on some widely used CVIs is provided in Section~\ref{sec:cvis}. However, these indices often provide different recommendations and unless a significant insight into the expected clusters is embedded, it becomes highly difficult to arrive at a common consensus. Moreover, these indices comment on the effectiveness of the clustering algorithm and not on the complete clustering framework and pre-processing steps. To address these issues, this paper proposes a novel objective validation strategy for the task of clustering residential electric demand profiles (see Section~\ref{sec:objectiveValidation}).

\subsection{Clustering Validity Indices (CVIs)}
\label{sec:cvis}
Although numerous CVIs have been used to validate the results for clustering residential load demand profiles~\cite{williams2013,zhou2017}, this paper focuses on a few which are most widely used in this context. For each household ($i$), the \textbf{Silhouette index} (SH)~\cite{rousseeuw1987} first measures the separation ($s_{i}$) by finding the mean pairwise distance between the household and the points from the nearest cluster of which the household is not a part of. Compactness ($c_i$) is then measured as the mean pairwise inter-cluster distance. Finally, the SH index is defined as: $\sum_{i=1}^{n} \frac{s_{i}-c_{i}}{n \cdot max(s_{i}, c_{i})}$, where $n$ is the total number of households which were clustered. The values of this index lie in the interval $[-1,\, 1]$, where a higher value represents better clustering.

The \textbf{Cali\'{n}ski-Harabasz} (CH) score~\cite{calinski1974}, also known as variance ratio criterion, defines separation by the inter-cluster dispersion (or the variance of cluster centers \textit{w.r.t.} the center of cluster centers weighted by the number of households in each cluster) and compactness as the within-cluster dispersion (or the average variance of points in a cluster \textit{w.r.t.} the cluster's center). On the other hand, \textbf{Dunn's Index} (DI)~\cite{dunn1973} defines separation as the minimum inter-cluster distance (i.e., the minimum distance between a pair of points that are not in the same cluster) and compactness as the maximum cluster diameter across all the clusters. Although both of these indices have their own definitions of separation and compactness, they are defined by the ratio of these two quantities. Additionally, a higher value resembles better clustering for both indices.

Similar to previous indices, the \textbf{Davies-Bouldin} (DB) index~\cite{davies1979} defines separation as inter-cluster distance ($s_{i,j}$ for cluster $i$ and $j$) and compactness as the average distance of each point within the cluster to its center ($c_{i}$ for cluster $i$). However, in this case, a different concept of similarity is introduced. Similarity between two clusters (say, $i$ and $j$) is the ratio of compactness and separation (i.e. $\frac{c_{i}+c_{j}}{s_{i,j}}$). For each cluster, the maximum similarity is computed, i.e. the similarity between the cluster and its most similar one. Finally, the average of these similarities is defined as the DB index. Since dislike clusters are better, lower values of this index are preferred.

Since FCM is considered as one of the clustering algorithms, the \textbf{Xie-Beni} (XB) index~\cite{xie1991} is also considered. This index defines the squared minimum inter-cluster distance as separation, and the mean squared distance between each point and the center of its assigned cluster as the compactness. Dividing compactness by separation defines the XB index. Similar to DB, a lower value of this index means better clustering. While the hard labels are directly interpreted as before, fuzzy centers are defined (as per equation~\ref{eq:fuzzyUpdates}) for this index where fuzzy clusters are formed, like in the case of the FCM clustering algorithm. Additionally, in the case of FCM, fuzziness ($u_{ij}^{m}$; for $i$\textsuperscript{th} cluster and $j$\textsuperscript{th} household) is also multiplied with squared distances while calculating the compactness of a cluster.

\subsection{Objective Validation Strategy}
\label{sec:objectiveValidation}

Our proposed objective evaluation strategy is based on a systematic treatment of the various residential electric demand profiles. Unlike in the case of CVIs (see Section~\ref{sec:cvis}), where the fitness of the clustering results was only defined by the input (dimensionally reduced features of the households) and output (cluster labels assigned to each household) arguments of the clustering algorithm, the proposed strategy takes a holistic view of the complete clustering process to comment on its fitness. Our proposed strategy utilizes the fact that during the pre-processing stage, the daily electricity consumption trend of each household $i$ (represented by a single feature vector of dimension $24/r$, see Algorithm~\ref{algo:preProcess}), is estimated from multiple days of data (say, $D_{i}^{days \times 24/r}$) collected by the smart-meters. In this study, we do so by calculating the median of the hourly (i.e. $r = 1$) consumption profiles for each household.

To validate the clustering results, using our objective validation strategy, we randomly shuffle the raw daily profiles for each household ($\vars{D}_{i}$), by rows, to obtain $\vars{D}'_{i}$ (of size $(days \times 24/r)$). We then make $\vars{p}$ equal partitions of $\vars{D}'_{i}$ and then use each such partition to create a daily load profile (using the median approach in this case) for the household. Hence, we will obtain $\vars{p}$ profiles for each household instead of just $1$. Now, assuming the following conditions, we can expect that the newly obtained $\vars{p}$ profiles are still representative of the household's consumption trend and should, at the very least, be clustered in the same cluster as the original non-partitioned profile.

\begin{itemize}
    \item $\vars{p} \lll\textrm{data available for anyone household}$, and
    \item the pre-processing algorithm can produce an accurate representation of the residential electric demand profiles.
\end{itemize}

\begin{algorithm}
\caption{Objective Validation Strategy}
\label{algo:eval}
%\DontPrintSemicolon % Some LaTeX compilers require you to use \dontprintsemicolon instead
\SetKwRepeat{Do}{do}{while}
$\vars{p} \gets \text{number of partitions for each household}$\\
$dist(\cdot, \cdot) \gets \text{function to calculate Euclidean distance}$\\
\Comment{Output from clustering algorithm}
$\textbf{\vars{labels}} \gets \text{labels assigned to each household}$\\
$\textbf{\vars{C}$_{(k \times d')}$} \gets \text{cluster centers of each cluster}$\\
\textbf{Initialize:}\linebreak
$\vars{nMatch}, \vars{nMisMatch}, \vars{counter} = 0$\;
$\vars{trials} \gets \text{number of times the process is repeated}$\\
\Repeat{$\vars{counter}<\vars{trials}$}{
\ForEach{household}{
    $\vars{D } \gets \text{data for each household (days$\times$24/r)}$\;
    $\vars{D'} \gets \text{randomly shuffled data by rows}$\;
    Make \vars{p} equal partitions from rows of \vars{D'}\;
    \Comment{Perform Pre-Processing steps}
    $\bm{M}_{\vars{p}}^{(\vars{p} \times (24/r))} \gets \text{new medians of } \vars{p} \text{ partitions}$\;
    $\bm{M}_{\vars{p}}^{\prime} \gets \ell_2\text{-Normalization(}\bm{M}_{\vars{p}}\text{, row-wise)}$\;
    \Comment{Do Dimensionality Reduction}
    $\bm{N}_{\vars{p}}^{(\vars{p} \times d')} \gets \text{dimReduce(}\bm{M}_{\vars{p}}^{\prime}\text{, } d'\text{)}$\;
    \ForEach{$partition \in \{\vars{1}\cdots\vars{p}\}$ \textbf{as} $part$}{
        \Comment{Find Closest Cluster}
        $\vars{CC} \gets \argmin_{i}(dist(\bm{N}_{\vars{p}}[part],\textbf{\vars{C}\,[i,\,:]}))$\;
        \eIf{$\vars{CC} == \textbf{\vars{labels}}\,[household]$}{
            $\vars{nMatch++}$\;
            }{
            $\vars{nMisMatch++}$\;
            }
        }
    }
    $\vars{counter++}$\;
}
\Comment{Compute average results per trial}
$\vars{avgMatches}=\vars{nMatch}/\vars{trials}$\;
$\vars{avgMisMatches}=\vars{nMisMatch}/\vars{trials}$\;
\KwResult{\vars{avgMatches} \& \vars{avgMisMatches}}
\end{algorithm}

The remaining pre-processing steps (performing $\ell_{2}$-Normalization of the median profiles) and dimensionality reduction techniques are generally deterministic, and therefore the similar steps can be applied to the $\vars{p}$ partitions. The output of the clustering algorithms is the corresponding cluster centers for all the clusters and a list of assigned labels to all the non-partitioned households. Thus, we can simply find the new label for each pre-processed and dimensionally reduced partition of the household by identifying the closest cluster center. If the cluster label assigned to a partitioned trend of a particular household remains the same as the original label that was assigned to the non-partitioned trend of the household, we call it a \textit{match}; otherwise, it is a \textit{mismatch}. This implies that a higher number of matches indicate better clustering. The results were averaged over $100$ trials to diminish the effect of biased orders during random shuffling. The complete framework of our objective validation strategy is shown in Algorithm~\ref{algo:eval}.

\section{Results}
\label{sec:results}
\subsection{Validation by CVIs}
\label{sec:resultsCVIs}
As discussed in Section~\ref{sec:cvis}, the values of Silhouette index (SH), Cali\'{n}ski-Harabasz (CH) score, Dunn's Index (DI), Davies-Bouldin (DB) index, and Xie-Beni (XB) index are computed for the $8$ clustering frameworks which were discussed in Section~\ref{sec:clusteringFramework}. These results are presented in Table~\ref{table:cviWithDimReduce}.

\begin{table}[htb!]
\small
\centering
\caption{PERFORMANCE EVALUATION AS PER THE STANDARD CLUSTER VALIDATION INDICES. THE $\uparrow$ (OR $\downarrow$) REPRESENTS WHETHER LARGER (OR SMALLER) VALUE INDICATES BETTER PERFORMANCE.
}%No partition is required for these validation indices (i.e. $p=1$).
\begin{tabular}{p{1.8cm}||c|c|c|c|c}
\hline
Methods & SH ($\uparrow$) & CH ($\uparrow$) & DI ($\uparrow$) & DB ($\downarrow$) & XB ($\downarrow$) \\ 
\hline\hline
PCA \& KMC\ & 0.21 & 11.69 & 0.35 & 1.04 & 0.40\\
FA \& KMC\ & 0.23 & 13.00 & \textbf{0.41} & \textbf{0.88} & 0.46\\
PCA \& SC\ & -0.06 & 3.87 & 0.18 & 1.35 & 1.74\\
FA \& SC\ & -0.01 & 4.28 & 0.18 & 1.42 & 2.34\\
PCA \& AC\ & 0.29 & 13.78 & 0.34 & 1.00 & 0.42\\
FA \& AC\ & \textbf{0.33} & \textbf{15.25} & 0.33 & 0.89 & 0.33\\
PCA \& FCM\ & 0.30 & 14.12 & 0.34 & 0.95 & 0.39\\
FA \& FCM\ & \textbf{0.33} & \textbf{15.25} & 0.33 & 0.89 & \textbf{0.32}\\
\hline
\end{tabular}
%\vspace{-0.5cm}
\label{table:cviWithDimReduce}
\end{table}
\vspace{0.5cm}

\begin{table*}[!hb]
\small
\centering
\caption{VALIDATION RESULTS OBTAINED BY THE PROPOSED STRATEGY FOR TWO DIFFERENT NUMBER OF PARTITIONS, $\vars{p}=2$ AND $\vars{p}=3$.}
%\begin{tabular}{ |p{0.4cm}||c|c|c|c|c| }
\begin{tabular}{ |lr||c|c|c|c|c|c| }
\hline
& Methods & \#Clusters & \#Total Cases & \#Avg. Matches & \#Avg. Mismatches & \%Matches & \%Mismatches \\ 
\hline\hline
\parbox[t]{3mm}{\multirow{8}{*}{\rotatebox[origin=c]{90}{\small $\bm{\vars{p} = 2}$}}} 
& PCA \& KMC\ & 7 & 54 & 41.52 & 12.48 & 76.89 & 23.11\\
& FA \& KMC\ & 7 & 54 & 21.50 & 32.50 & 39.81 & 60.19 \\
& PCA \& SC\ & 9 & 54 & 9.86 & 44.14 & 18.26 & 81.74\\
& FA \& SC\ & 7 & 54 & 12.55 & 41.45 & 23.24 & 76.76\\
& PCA \& AC\ & 4 & 54 & 45.43 & 8.57 & 84.13 & 15.87\\
& FA \& AC\ & 4 & 54 & 32.28 & 21.72 & 59.78 & 40.22\\
& PCA \& FCM\ & 4 & 54 & 47.39 & 6.61 & \textbf{87.76} & 12.24\\
& FA \& FCM\ & 4 & 54 & 30.40 & 23.60 & 56.30 & 43.70\\
\hline\hline
\parbox[t]{3mm}{\multirow{8}{*}{\rotatebox[origin=c]{90}{\small $\bm{\vars{p} = 3}$}}} 
& PCA \& KMC\ & 7 & 81 & 54.23 & 26.77 & 66.95 & 33.05\\
& FA \& KMC\ & 7 & 81 & 29.45 & 51.55 & 36.36 & 63.64\\
& PCA \& SC\ & 9 & 81 & 12.97 & 68.03 & 16.01 & 83.99\\
& FA \& SC\ & 7 & 81 & 16.32 & 64.68 & 20.15 & 79.85\\
& PCA \& AC\ & 4 & 81 & 65.64 & 15.36 & 81.04 & 18.96\\
& FA \& AC\ & 4 & 81 & 42.78 & 38.22 & 52.81 & 47.19\\
& PCA \& FCM\ & 4 & 81 & 67.30 & 13.70 & \textbf{83.09} & 16.91\\
& FA \& FCM\ & 4 & 81 & 43.59 & 37.41 & 53.81 & 46.19\\
\hline
\end{tabular}
%\vspace{-0.5cm}
\label{table:scoretable}
\end{table*}

As speculated before in Section~\ref{sec:valCFs}, it can be noticed that the different CVIs give different recommendations while comparing the clustering frameworks. While DI and DB indices favor the `FA \& KMC' framework, the XB index favours the `FA \& FCM' framework, and the other indices recommend `FA \& AC' alongside `FA \& FCM'. It must be noted here that since the SH, CH, DI, and DB indices are not meant to deal with fuzzy labels directly, the fuzzy labels which were generated by the FCM algorithm were converted into hard labels by assigning each household to the cluster with which it exhibits maximum membership.

\subsection{Validation by the Proposed Strategy}
\label{sec:resultsProposedStrategy}
To obtain a more holistic recommendation by analyzing the whole clustering framework, a novel objective validation strategy is defined in Section~\ref{sec:objectiveValidation}. Due to the limited data availability in the used dataset, only small partitions (i.e., $\vars{p} = 2$ and $\vars{p} = 3$) are considered in the experiments. This was done in order to comply with the assumptions made while describing the strategy. Table~\ref{table:scoretable} presents the results obtained by our objective validation strategy. As inferred from Algorithm~\ref{algo:eval}, the total number of cases is given by the number of partitions that were made for each household ($\vars{p}$) times the number of households present in the original dataset ($n$). This is noted in the third column of the table. While the number of average matches (\#Avg. Matches) and mismatches (\#Avg. Mismatches) were provided directly by Algorithm~\ref{algo:eval}, percent matches (\%Matches) and mismatches (\%Mismatches) were also computed to better compare the final results. A higher value of percent matches represents a better and more stable clustering framework. As the number of partitions increases (i.e. moving from $\vars{p} = 2$ to $\vars{p} = 3$), the variability in the input increases and the hence the percentage of matches decreases. Furthermore, it is perspicuous from the table that the proposed objective validation strategy gives a clear recommendation for the `PCA \& FCM' framework. \textit{Note:} Similar to the case of SH, CH, DI, and DB indices, fuzzy labels, generated by the FCM algorithm, are also transformed into hard labels using the same procedure.

\subsection{Discussion and Comparative Analysis}
\label{sec:discussion}
It can be noted from the results described in Table~\ref{table:cviWithDimReduce} and Table~\ref{table:scoretable} that the proposed objective validation strategy favors PCA variants as opposed to FA, whereas the CVIs are more inclined towards FA. This can be easily attributed to the fact that the CVIs do not consider the dimensionality reduction step while making their recommendations. On the other hand, the proposed objective validation strategy not only considers the results of the pre-processing and dimensionality reduction steps but follows the complete process while calculating the number of matches and mismatches (see Algorithm~\ref{algo:eval}),  based on which it makes its recommendations. To support this explanation, the CVIs were recalculated in an unorthodox manner by using the original (i.e. non-dimensionally reduced) features of the households instead of the ones obtained after the dimensionality reduction step. These results are presented in Table~\ref{table:cviWithoutDimReduce}. When the complete framework is considered, even the CVIs tend to prefer the PCA variants over the FA ones. Furthermore, $4$ out of $5$ CVIs recommend the `PCA \& FCM' clustering framework, reinforcing the results obtained by our proposed method.

\begin{table}[htb!]
\small
\centering
\caption{RECALCULATED CVI VALUES BY CONSIDERING NON-DIMENSIONALLY REDUCED PROFILES OF HOUSEHOLDS INSTEAD OF THE DIMENSIONALLY REDUCED ONES AS USED IN TABLE~\ref{table:cviWithDimReduce}. THE $\uparrow$ (OR $\downarrow$) REPRESENTS WHETHER LARGER (OR SMALLER) VALUE INDICATES BETTER PERFORMANCE.}
\begin{tabular}{p{1.8cm}||c|c|c|c|c}
\hline
Methods & SH ($\uparrow$) & CH ($\uparrow$) & DI ($\uparrow$) & DB ($\downarrow$) & XB ($\downarrow$) \\ 
\hline\hline
PCA \& KMC\ & 0.19 & 10.18 & 0.38 & 1.12 & 0.45\\
FA \& KMC\ & 0.18 & 9.95 & \textbf{0.40} & 1.03 & 0.60\\
PCA \& SC\ & -0.04 & 3.68 & 0.20 & 1.37 & 1.68\\
FA \& SC\ & -0.01 & 3.55 & 0.20 & 1.55 & 1.87\\
PCA \& AC\ & 0.27 & 12.52 & 0.37 & 1.06 & 0.46\\
FA \& AC\ & 0.27 & 12.52 & 0.37 & 1.06 & 0.46\\
PCA \& FCM\ & \textbf{0.28} & \textbf{12.80} & 0.37 & \textbf{1.02} & \textbf{0.43}\\
FA \& FCM\ & 0.27 & 12.52 & 0.37 & 1.06 & 0.44\\
\hline
\end{tabular}
%\vspace{-0.5cm}
\label{table:cviWithoutDimReduce}
\end{table}

Moreover, it should be noted that the three frameworks `PCA \& AC', `FA \& AC', and `FA \& FCM' generate identical clustering results, i.e. they assign households to the clusters in an identical manner. However, the validation results are identical for these frameworks only in Table~\ref{table:cviWithoutDimReduce}. The underlying reason is that the CVIs consider only two parameters, i.e. the features of the households and their labels. While calculating CVIs for Table~\ref{table:cviWithoutDimReduce}, these two parameters were the same, hence the identical results. Whereas when CVIs were calculated for Table~\ref{table:cviWithDimReduce}, the reduced features were used which were different for `PCA' and `FA', hence the results are different for the `PCA \& AC' framework but are still identical for the other two. However, the results are completely different for our proposed method since it considers the whole process and not only the input and output arguments of the clustering framework. This further strengthens the importance of the proposed validation strategy, especially in the cases where the designed optimal clustering framework is required to be processed again in order to cluster the newly obtained data for the households~\cite{mcloughlin2015,lin2017}.

\section{Conclusion \& Future Work}
\label{sec:conclusion}
Data collected from smart meters at sufficient and high resolutions is crucial for developing various insights about customer consumption patterns. Clustering residential load profiles is an important step to perform such analysis. This paper defines a generalized framework to cluster these profiles. While many algorithms can easily be incorporated into the defined framework, choosing the most appropriate one is often the most challenging task. To this end, the paper proposes a novel objective validation strategy that not only considers the clustering results but takes the complete clustering framework into the account.

The proposed validation strategy is further compared with the standard CVIs and it is shown that the proposed scheme has several advantages over the standard CVIs. While standard CVIs primarily focus only on the intrinsic properties of the clustering algorithm, the proposed strategy takes a holistic look at the complete clustering framework as well as the pre-processing steps to make its recommendations. Due to this fact, the proposed strategy makes better, unbiased, and uniform recommendations as compared to the standard CVIs making it a preferred candidate to compare the results of different clustering frameworks.

Furthermore, it was noted that different CVIs provide different recommendations with the standard approach (Table~\ref{table:cviWithDimReduce}). Whereas the optimal clustering framework for the given dataset (i.e. PCA \& FCM) was not recommended by any CVI. The paper identifies the issue as the lack of a holistic view with the CVI approach. This limitation, however, is lifted with the proposed objective validation strategy. The proposed strategy re-evaluates the whole clustering framework (including some pre-processing steps) to generate a metric called \%Matches. A higher value of this metric indicates better clustering. For PCA \& FCM framework, we obtained a high value of $87\%$ Matches. The superiority is further cross-validated by re-calculating CVI values over the non-dimensionally reduced profiles (in a way, forcefully asking them to get a broader outlook over the whole framework).

On the downside, however, the proposed strategy is not as generic as the standard CVIs. This is because the former utilizes the fact that the load consumption data, which is available for different days, is reduced to a single representative daily consumption profile by computing the median in the pre-processing stage. Since it is not always the case in unsupervised clustering tasks, the proposed strategy cannot be used off the shelf everywhere just like the standard CVIs. Nevertheless, the proposed strategy is a great tool to determine the most optimal clustering framework when it comes to the task of clustering residential electric load demand profiles where daily load demand profiles are generally estimated from the consumption data gathered for multiple days.

The ability of the proposed objective validation strategy to identify most optimal clustering framework shows promise and superiority to standard CVIs, specifically for the task of clustering household electric load demand profiles. Moreover, no assumption is made with regard to the datasets (i.e. the size of the dataset or the case study it was obtained from). Hence, the proposed strategy is easily reproducible. Furthermore, claiming to give unbiased results while having a holistic view of the whole framework, it practically removes the requirement of manual inspection of the generated clusters (or subjective validation), thereby making the process of optimal clustering scalable for large number of households. In the future, we aim to cross-validate these aspects by performing experiments with more datasets obtained from different case studies and having a larger number of households.

\bibliographystyle{IEEEtran.bst}


\vspace{-1.6em}
\begin{IEEEbiography}[{\includegraphics[width=1in,height=1.25in,clip,keepaspectratio]{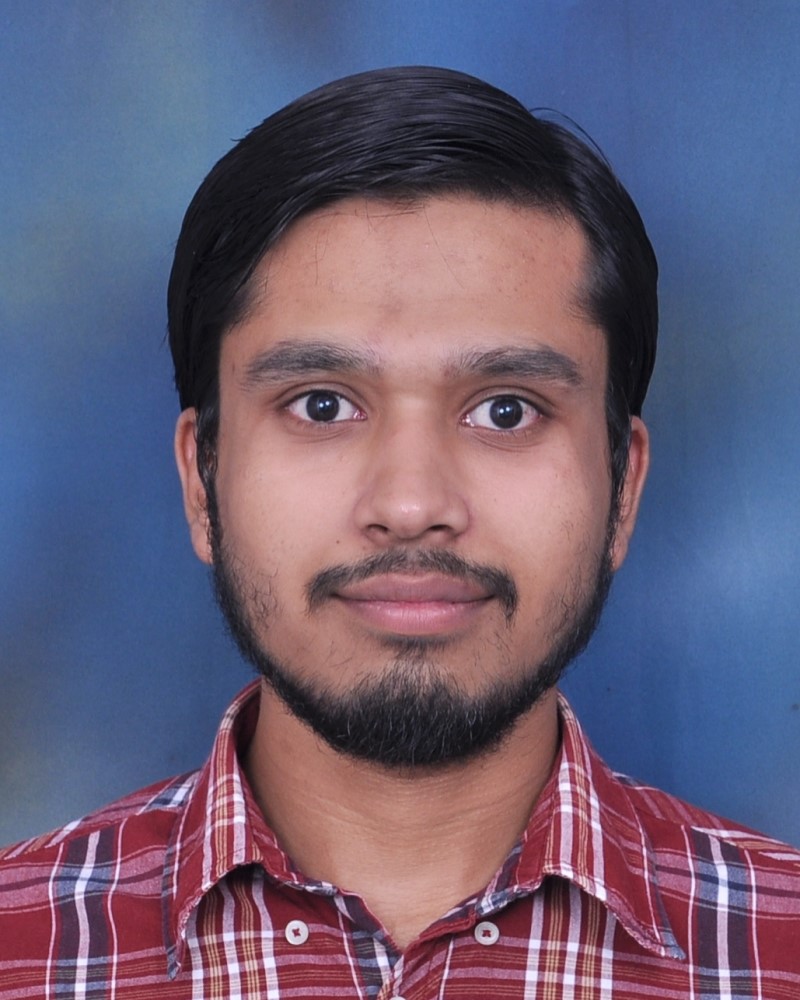}}]{Mayank Jain} is a PhD scholar at the School of Computer Science, University College Dublin since $2020$. His doctoral thesis is focused on facilitating the inclusion of solar power in energy grids. At present, he is also affiliated with ADAPT SFI Research Centre, Dublin, Ireland. In $2015$, he completed his Bachelors in IT and Mathematical Innovations with minor specialization in Electronics and Robotics from the Cluster Innovation Centre, University of Delhi. Afterwards, he founded a start-up aimed at promoting STEM education in Indian schools ($2015-18$). With prestigious Scottish Saltire scholarship award from the government of Scotland, he resumed his academic career with the Masters in Artificial Intelligence from the University of Edinburgh ($2018-19$).
\end{IEEEbiography}
\vspace{-1.6em}
\begin{IEEEbiography}[{\includegraphics[width=1in,height=1.25in,clip,keepaspectratio]{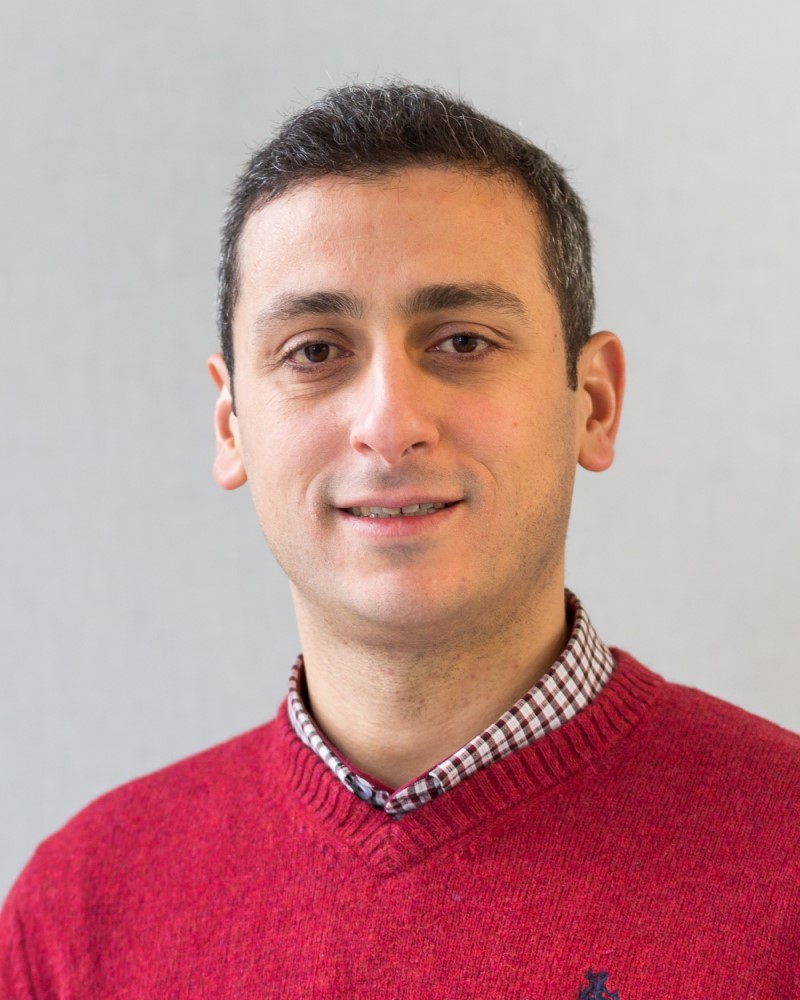}}]{Tarek AlSkaif} is an Assistant Professor at the Information Technology group (INF), Wageningen University \& Research (WUR). He received the Ph.D. degree (Cum-laude) from Universitat Polit\`{e}cnica de Catalunya (UPC), Statistical Analysis of Networks and Systems (SANS) research group, Barcelona, Spain in $2016$ and obtained the B.Sc. degree and the M.Sc. degree in Computer and Electrical Engineering from Damascus University, Syria in $2007$ and $2011$, respectively. Between $2016 - 2020$, he was a postdoc at the Energy \& Resources group at the Copernicus Institute of Sustainable Development, Utrecht University (UU), The Netherlands. His research interests span over different problems in smart grids, including smart energy systems design, local energy communities, smart integration of distributed energy resources, with an emphasis on using ICT platforms, mathematical modeling \& optimization and artificial intelligence. Tarek is the (co)author of over 50 peer-reviewed scientific articles, a regular reviewer and guest-editor in top international journals and conferences in the area of smart grids and energy informatics. 
\end{IEEEbiography}
\vspace{-1.6em}
\begin{IEEEbiography}[{\includegraphics[width=1in,height=1.25in,clip,keepaspectratio]{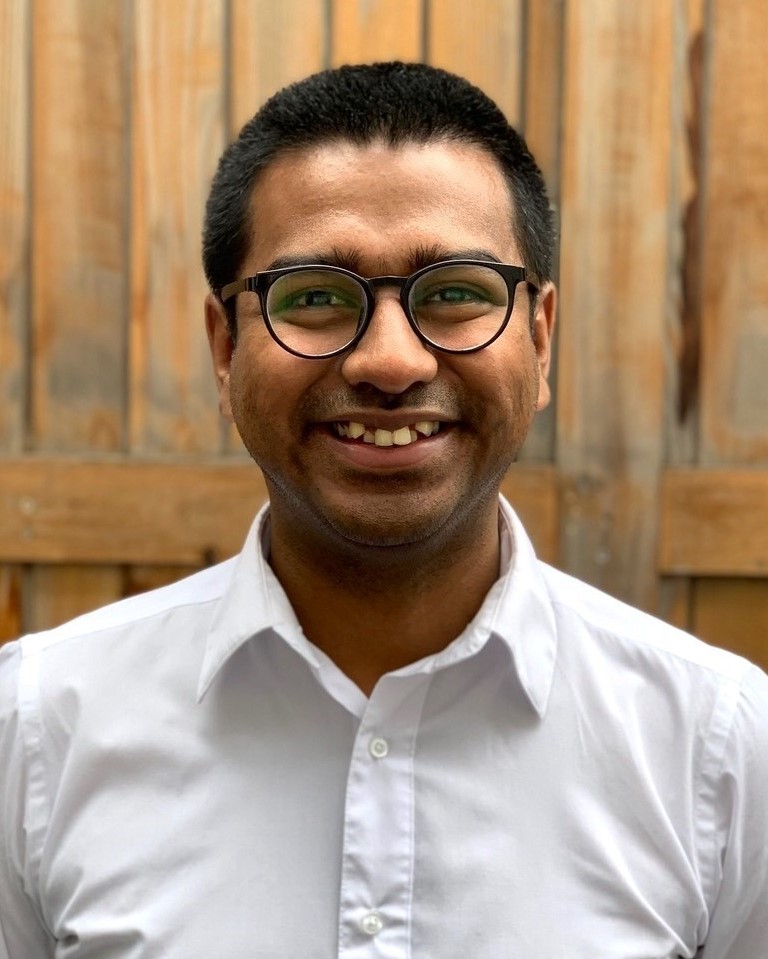}}]
{Soumyabrata Dev} (S'$09$-M'$17$) is an Assistant Professor at the School of Computer Science, University College Dublin. Prior to that, he was a postdoctoral researcher at ADAPT SFI Research Centre, Dublin, Ireland. He obtained his PhD from Nanyang Technological University (NTU) Singapore, in $2017$. From Aug-Dec $2015$, he was a visiting student at Audiovisual Communication Laboratory (LCAV), \'{E}cole Polytechnique F\'{e}d\'{e}rale de Lausanne (EPFL), Switzerland. He graduated summa cum laude from National Institute of Technology Silchar, India with a B.Tech. in $2010$. He is an IEEE member and has published over $60$ papers. His research interests include remote sensing, statistical image processing, machine learning, and deep learning.
\end{IEEEbiography}

%\balance

\end{document}